\documentclass[sort&compress
    ,final            
  ]
  {aipproc}

\usepackage{amsmath}
\usepackage{amssymb}

\layoutstyle{8x11double}

\newcommand{\Porb}{P_{\mathrm{orb}}}
\newcommand{\Pspin}{P_{\mathrm{spin}}}
\newcommand{\Mdot}{\dot{M}}

\newcommand{\msun}{M_{\odot}}

\begin{document}

\title{The Connection Between Low-Mass X-ray Binaries and (Millisecond) Pulsars: A Binary Evolution Perspective}


\author{Christopher J. Deloye}{
  address={Dearborn Observatory, Northwestern University, 2131 Tech Drive, Evanston, IL, 60208, cjdeloye@northwestern.edu}
}

\begin{abstract}
I review the evolutionary connection between low-mass X-ray binaries (LMXBs) and pulsars with binary companions (bPSRs) from a stellar binary evolution perspective.  I focus on the evolution of stellar binaries with end-states consisting of a pulsar with a low-mass ($<1.0 \msun$) companion, starting at the point the companion's progenitor first initiates mass transfer onto the neutron star.  Whether this mass transfer is stable and the physics driving ongoing mass transfer partitions the phase space of the companions's initial mass and initial orbital period into five regions. The qualitative nature of the mass-transfer process and the binary's final end-state differ between systems in each region; four of these regions each produce a particular class of LMXBs.  I compare the theoretical expectations to the populations of galactic field LMXBs with companion-mass constraints and field bPSRs.  I show that the population of accreting millisecond pulsars are all identified with only two of the four LMXB classes and that these systems do not have readily identifiable progeny in the bPSR population.  I discuss which sub-populations of bPSRs can be explained by binary evolution theory and those that currently are not.  Finally I discuss some outstanding questions in this field.
\end{abstract}


\maketitle


\section{Introduction}
Since the discovery of the class prototype \citep{backer82}, there has been a posited evolutionary connection between millisecond radio pulsars (MSPs) and low-mass X-ray binaries (LMXBs)---mass transferring binaries with a neutron star (NS) accretor and donor companion with a mass  $M_2 \lesssim 1 \msun$ \citep{alpar82}.  The central idea behind this connection is that LMXBs can provide the long-lived phase  ($\sim \mathrm{Gyr}$) of moderate mass transfer rates ($\Mdot \lesssim \Mdot_{\mathrm{Edd}} \approx 10^{-8} \msun\,\mathrm{yr}^{-1}$, where $ \Mdot_{\mathrm{Edd}}$ is the Eddington mass-transfer rate) thought necessary to spin-up the NS to spin periods of $\Pspin < 10 \mathrm{ms}$ as observed in the MSP population. 

\begin{figure}
  \includegraphics[height=.425\textheight]{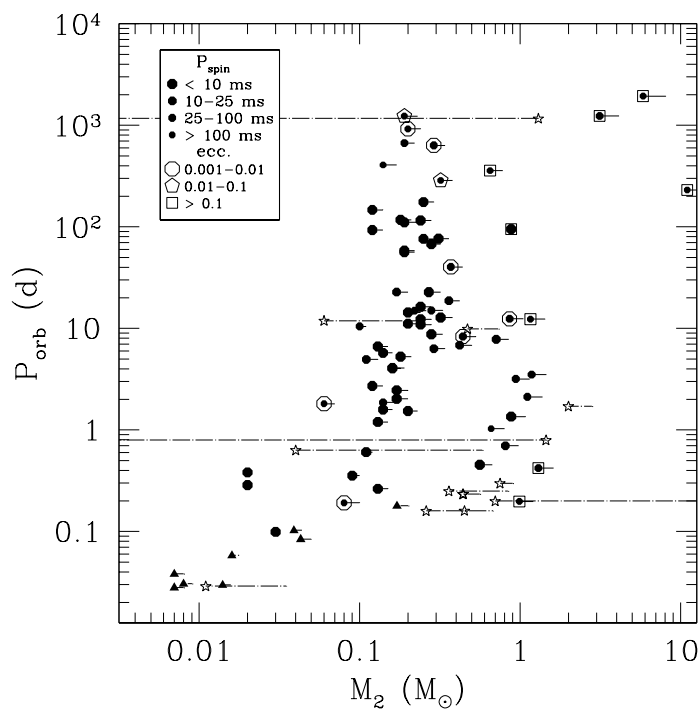}
  \caption{The galactic field population of radio pulsars with binary companions (filled circles) in the $M_2$-$\Porb$ plane. The data are from the ATNF catalog \citep{manchester05}. Plotted are each system's minimum $M_2$, with horizontal lines extending to the system's median $M_2$ (assuming $i$ is randomly distributed). Filled  circle size indicates each pulsar's $\Pspin$ (see plot legend). Symbols circumscribing a filled circle indicates the binary's eccentricity (again see plot legend).  For comparison, LMXBs with independent $M_2$ estimates are plotted with open stars and filled triangles, the latter indicating the minimum $M_2$ for \emph{accreting MSP} systems.  For the open stars, the dash-dotted lines indicate the \emph{total} estimated $M_2$ range in each system. For the accreting MSPs, the dash-dotted lines extend to the median $M_2$.  This sample of LMXB systems represents a union of systems with $M_2$ estimates in the \citet{ritter98} catalog and a targeted literature search on LMXB systems where $\Pspin$ has been determined.  Thus, this plot likely does not present in total the current census of LXMBs with $M_2$ estimates. }
  \label{fig:bPSRs_sys}
\end{figure}

Observational evidence for millisecond variability in LMXBs has been growing steadily in the form of detections of kilohertz quasi-periodic oscillations (kHz QPOs), X-ray burst oscillations, and accretion-powered oscillations \citep[see][]{chakrabarty05}. In terms of support for the LMXB-MSP connection, pride of place has been given to the accretion-powered pulsations systems (also known as accreting MSPs) since the pulsation period in these systems are identifiable directly with $\Pspin$ \citep{chakrabarty05}.   However, recent work has also shown that $\Pspin$ is of order the oscillation periods in the kHz QPO and X-ray burst oscillation sources \citep{chakrabarty03,strohmayer03,wijnands03,linares05}, establishing that these systems also harbor a rapidly rotating NS.  Thus, it is now well established that NSs in LMXBs can be spinning rapidly enough to produce radio MSPs once the LMXB phase ends.

However, understanding fully the evolutionary connection between LMXBs and MSPs requires not only explaining the NS's spin evolution but also accounting for other properties seen in the radio MSP population.  In particular, this includes the distribution in orbital period, $\Porb$ of MSPs that retain a remnant binary companion, how this remnant's $M_2$ correlates with $\Porb$, the distribution of binary eccentricity, $e$, and the production of \emph{isolated} MSPs.  Indeed, the ideal test of accretion-torque theory would be accomplished by understanding the evolution in the LMXB-phase well enough to correlate final $\Pspin$ with these other quantities.  This is a rather ambitious goal since the $\Pspin$ evolution depends not only on the secular $\Mdot$ evolution but also on the efficiency with which matter accretes onto the NS, whether accretion onto the NS occurs sporadically due to disk instabilities, how $\Pspin$ evolves during mass-transfer outbursts and when unstable disks are in quiescent phases, and how each LMXB transitions into an MSP system. 

Turning from where one would like to be to where we are now, my goal for this contribution is to approach the LMXB-MSP connection from the vantage point of stellar binary evolution theory.  To do so, I will expand the view somewhat and review our understanding of NS-main sequence (MS) binaries whose evolutionary endpoints are NSs with a low-mass ($M_2 \lesssim 1.0 \msun$) binary companion. In doing so, the focus of the discussion will shift from solely the MSP population to making connections between NS-MS binaries and various populations of radio pulsars in binaries, bPSRs (one would like to simply say ``binary pulsars'' here, but the discovery of \emph{the} Binary Pulsar \citep{burgay03} has lead to this term often causing confusion; I'll leave it to the reader to decide whether the scientific windfall from this system compensates sufficiently for necessitating such unwieldy terminology).   

The starting point for this discussion is Figure \ref{fig:bPSRs_sys}, which shows the location of galactic field bPSRs in the $\Porb$-$M_2$ plane by the filled circles \citep{manchester05}. Only field sources are included so as to compare theory to a sample of bPSRs whose properties have not been influenced by dynamic interactions.  The filled circles indicate each bPSR's minimum-$M_2$; the horizontal lines extend to each system's median $M_2$ (corresponding to a binary inclination of $i = 60^\circ$).  Spin period and $e$ are encoded via filled circle size and circumscribed symbols, respectively. Filled triangles show the same information for accreting MSP systems \citep{chakrabarty98,galloway02,galloway05,kaaret06,krimm07,markwardt02,markwardt03}.  Open stars indicate other field LMXB systems with $M_2$ determinations \citep{bhattachar06,casares06,cominsky89,cornelisse07,finger96,heinz01,hinkle06,jonker01,parmer86,pearson06,reynolds97}; for these latter systems, the dashed-dotted horizontal lines show the estimated range of $M_2$ (not its median value). 
 
In the following, I'll compare the predictions of binary evolution theory for how systems evolve in this $\Porb$-$M_2$ plane to the location of systems in Fig. \ref{fig:bPSRs_sys}. While this will allow tentative positive identifications of the evolutionary connections discussed above,  it will also serve to highlight sub-populations of bPSRs whose formation is \emph{not} currently explained by binary evolution theory. I'll discuss the evolution of NS-MS binaries, focusing on how initial conditions lead to four different classes of X-ray binaries.  I'll compare between this theory and the observations, pointing out where agreement between the two is better and worse. Finally, I'll close by discussing several open questions related to the evolution of LMXBs and bPSR formation.  For other discussions and reviews of the bPSR population see \citep{lorimer05} and \citep{phinney94}. Also see \citep{flamb05} for a conference proceeding that reviews NS spin evolution under accretion.

\section{Four Classes of LMXBs}
The subsequent evolution of a binary systems that begins mass transfer depends on several factors: the response of the donor's radius, $R_2$, to mass loss, whether the mass transfer is stable or not, and the mechanism that drives continued mass transfer when it is stable.  The $R_2$ response is usually characterized by the quantity $\xi_2$:
\begin{equation}
\xi_2 \equiv \frac{d\,\ln R_2}{d\,\ln M_2}\,.
\end{equation}
Whether mass transfer is stable depends on the relative evolution of $R_2$ and the donor's Roche radius, $R_L$, where $R_2 \approx R_L$ is required for mass transfer to take place.  Stable mass transfer occurs when
\begin{equation}
\xi_2 > \xi_L \,,
\end{equation}
where
\begin{equation}
\xi_L \equiv \frac{d\,\ln R_L}{d\,\ln M_2} \,,
\end{equation}
and depends on the binary's mass ratio ($q = M_2/M_{\mathrm{NS}}$) and whether mass loss is conservative or not \citep[see, e.g., ][]{tauris99}.  In the conservative case, $\xi_L = -2 ( 5/6 - q)$. In words, the stability criteria requires that transferring mass acts to reduce the extent the donor fills its Roche lobe.  Thus when mass transfer is stable, some external driver must act to maintain the condition $R_2 \approx R_L$.  Examples of such drivers include orbital angular momentum ($J$) loss mechanisms (which decrease $R_L$) or the donor's internal evolution (which can increase $R_2$, e.g., during evolution up one of the giant branches).  

\begin{figure}
  \includegraphics[height=.425\textheight]{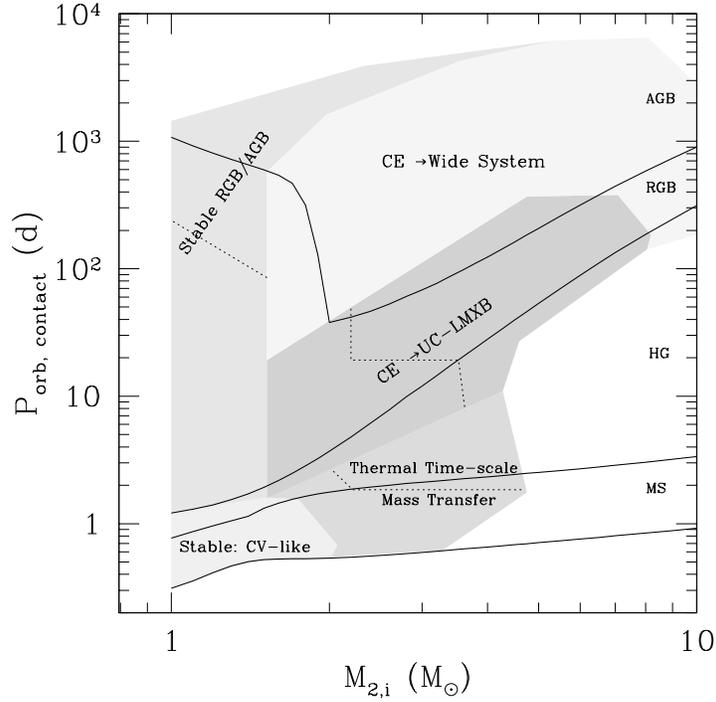}
  \caption{The phase space of $\Porb$ at onset of mass transfer versus initial main-sequence companion mass, $M_{2,i}$, of NS-MS binaries leading to various outcomes, indicated by labeled gray-scale shaded regions as calculated with the BSE code \citep{hurley02}.  These regions are meant to be indicative as the detailed boundaries depend on the detailed treatment of stellar and binary evolution (e.g., parameterizations of wind mass loss and common envelope). The heavy solid lines indicate boundaries between donor evolutionary states as calculated with the SSE code \citep{hurley00}. The initial stability of mass transfer and the physical process that drive ongoing mass transfer underlie this phase-space partitioning.  The dotted lines in several of the regions indicate the boundary between NS-companions that leave behind a He WD (lower-left portion of these regions) and either hybrid or C/O WDs (upper-right).  }
  \label{fig:lmxb_ips}
\end{figure}

\begin{figure}
  \includegraphics[height=.35\textheight]{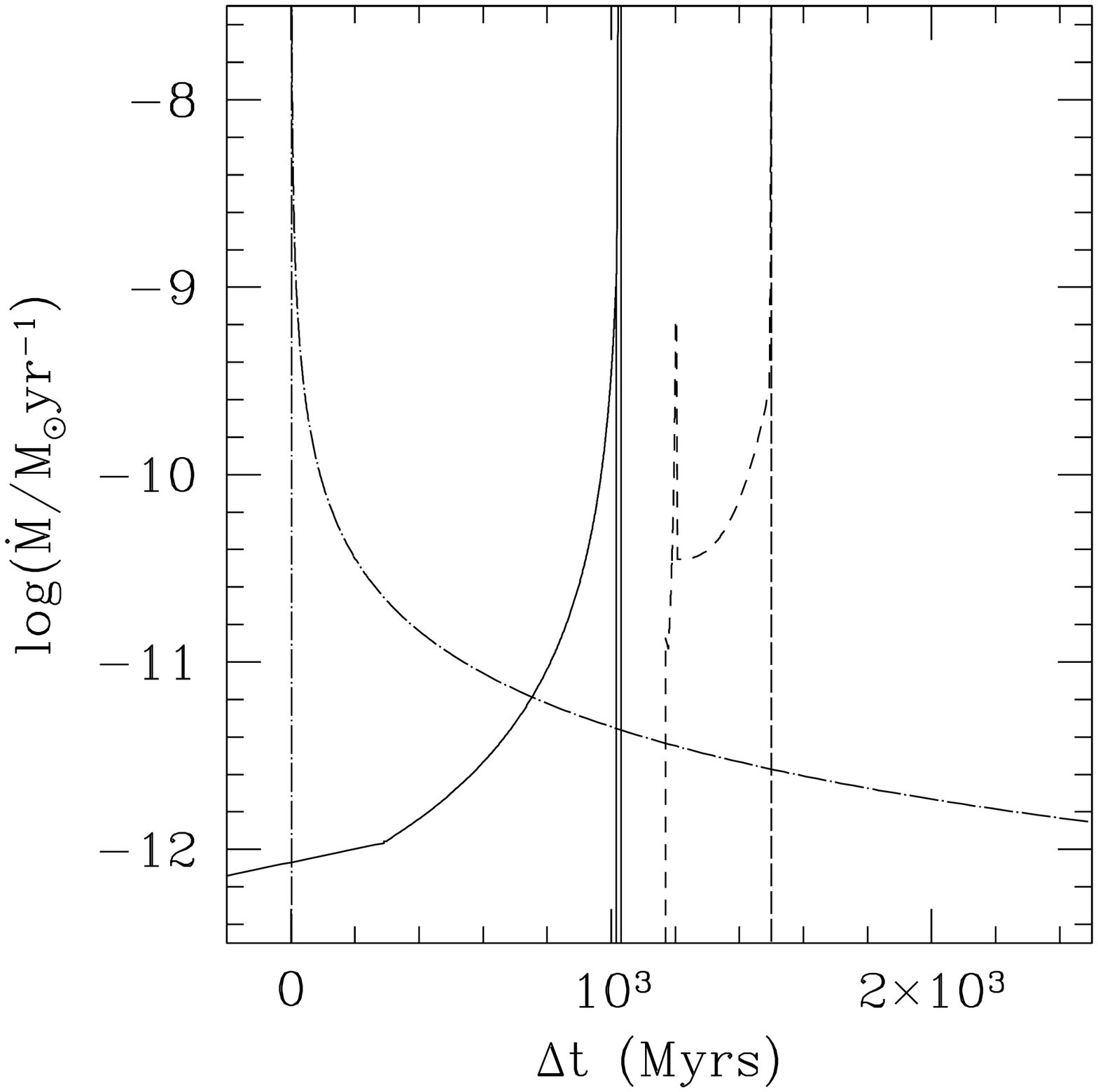}
  \caption{Representative examples of the $\Mdot$-evolution for LMXBs with giant-branch donors (solid line), those undergoing thermal time-scale mass transfer (dashed lines), and ultracompact LMXB systems (dash-dotted lines).  While LMXBs in the two former cases all experience a very rapid shut-off in mass-transfer at the end of the LMXB phase, ultracompact LMXBs (as well as CV-like systems) have a very gradual decay in $\Mdot$.  The RGB/AGB and thermal time-scale $\Mdot$ evolution were calculated with BSE \citep{hurley02}.  The UC-LMXB evolution was calculated using the code developed for \citet{deloye07}.} 
\label{fig:mdotcomp}
\end{figure}

The value of $\xi_2$ depends on the details of any given situation, but it has two limiting cases. Mass loss perturbs the donor away from dynamic and thermal equilibrium (the latter corresponding heuristically to the state where luminosity is constant throughout the donor's envelope).  The donor's response is to adjust towards a new dynamic and thermal equilibrium at its new mass. If mass loss proceeds on a time-scale ($\sim M_2/\Mdot$) longer than that of this thermal readjustment, $R_2$ evolves along the sequence set by these equilibrium states. This sets the limiting case for low $\Mdot$-values: $\xi_2 = \xi_{\mathrm{eq}}$.  In the opposite limit, when mass loss proceeds on a time-scale much shorter than the thermal readjustment time, the donor only has time to respond dynamically to  mass loss.  Then, mass elements in the donor evolve adiabatically and this sets the response in the high $\Mdot$-limit: $\xi_2 \equiv \xi_{\mathrm{ad}}$. In the adiabatic limit, the entropy profile of the donor's envelope determines whether the donor expands or contracts: convective envelopes lead to $\xi_2 \lesssim 0$, radiative envelopes produce $\xi_2 > 0$ \citep[see,][for discussions of this]{faulkner76,kalogera96,deloye07}.

How all of this plays out in setting the evolution of NS-accretor binaries when mass transfer begins is summarized in Figure \ref{fig:lmxb_ips}, which is a plot of the contact-$\Porb$ versus initial MS mass plane.  When  $R_2 = R_L$, the donor's mean density determines $\Porb$, so in this plot there is a direct correspondence between the donor's evolutionary state (which sets $R_2$) and the contact-$\Porb$.  This correspondence is shown with the heavy solid lines in Fig. \ref{fig:lmxb_ips} indicating boundaries between various donor evolutionary states: MS, Hertzsprung gap (HG), red-giant branch (RGB), and asymptotic-giant branch (AGB) phases. The five different shaded regions separate initial conditions based on modes and outcomes of mass transfer in these binaries as summarized below \citep[see also][]{kalogera96}.  The boundaries of these regions are meant to be indicative and approximate since they can depend on the detailed treatment of both stellar and binary evolution and on the NS's mass, $M_{NS}$ (here assumed to be $1.4 \msun$.).

\begin{description}
\item[Stable: Cataclysmic Variable (CV)-like] This region consists of low-mass donors making contact before evolving significantly up the RGB.  The smaller $q$-values in such systems produce stable mass transfer and at these shorter contact $\Porb$, $J$-loss mechanisms \citep[specifically magnetic braking][]{spruit83} dominate over the donor's nuclear evolution \citep{pylyser88} and this provides the driver for continued mass loss. The $\Mdot$ rates are low enough that $\xi_2 \approx \xi_{\mathrm{eq}} > 0$, resulting in contraction and evolution to shorter $\Porb$ (at least initially).

As systems evolve to shorter $\Porb$ and $M_2$ decreases, the donor eventually becomes fully convective. At this point magnetic braking is thought to stop operating \citep{spruit83,rappaport83} and the slower $J$-loss mechanism of gravity-wave (GW) emission becomes the dominant mass-transfer driver.  Further along, $M_2$ decreases sufficiently to shut-off nuclear burning, increasing the donor's thermal time-scale and producing a transition from an equilibrium to adiabatic response to mass loss.  The now fully convective donor has $\xi_{\mathrm{ad}} \lesssim 0$, and the $\Porb$-evolution by this time has reversed sign. The value of the $\Porb$-minimum is smaller for systems making contact as more evolved MS/HG objects and can reach into the $\Porb$ range of ultracompact LMXBs (see below). 

Once begun, the mass-transferring LMXB phase can persist upwards of 10 Gyr. 
Relevant modelling papers include \citep{pylyser88,podsiadlowski02,nelson04,vandersluys05}. 
\end{description}
\begin{description}
\item[Stable: RGB/AGB] This region consists of low-mass systems that initiate mass transfer on one of the giant branches. Again, $q$ is small leading to stable mass transfer (the tail of systems with large $M_{2,i}$ near the tip of the AGB  make contact after significant amounts of mass has been mass lost in winds).  Continued mass transfer is driven by the donor's nuclear evolution and resulting expansion during the giant phases.  This drives the system to wider orbital separations and, typically, longer $\Porb$.

The duration of the LMXB phase depends on $\Porb$ at contact, ranging from $\sim 1 \mathrm{Gyr}$ at $P_{\mathrm{orb,contact}} \sim 1 \mathrm{d}$ to $\sim 1 \mathrm{Myr}$ for initially wide systems.  Since $\Mdot$ is set by the donor's nuclear evolution rate, $\Mdot$ increases with $\Porb$ and is typically super-Eddington for systems with $P_{\mathrm{orb,contact}} \gtrsim 10 \mathrm{d}$.  In these cases, much of the transferred mass is likely ejected from the system \citep{kalogera96}, resulting in less efficient NS spin-up. 

Finally, the core mass-radius relation for giant branch stars with degenerate cores produces a relationship between the remnant $M_2$ and final $\Porb$ as discussed by \citep{rappaport95,tauris96,refsdal71,webbink83,joss87}. Additional, relevant modelling papers include \citep{tauris99,nelson04}.
\end{description}
\begin{description}
\item[Thermal Time-scale Mass Transfer] This mode of mass-transfer occurs for intermediate mass donors ($2.0 \lesssim M_{2,i} \lesssim 4.0 \msun$) that make contact before reaching the RGB.  The larger $q$ here lead to $\xi_L$ such that $\xi_{\mathrm{eq}} < \xi_L < \xi_{\mathrm{ad}}$: that is, $R_L$ evolves at a rate intermediate between the donor's equilibrium and adiabatic response ( $\xi_{\mathrm{ad}} \gg 1$ due to these donor's radiative envelopes). Thus, in trying to attain its new equilibrium $R_2$ by thermal adjustment, the donor is forced to fill its Roche-lobe.  This drives the continued mass-transfer in these systems.  The $\Porb$-evolution is determined by $\xi_L$: $\Porb$ initially decreases but as $q$ is reduced by mass transfer, $\xi_L$ decreases sufficiently to begin outward $\Porb$-evolution. The $\Mdot$ rates produced here can be very high, with peak values typically $\gg \Mdot_{\mathrm{Edd}}$.

The duration of the LMXB phase and system endpoints correlate with the companion's remnant.  Systems leaving behind a He WD have LMXB phases lasting between $100 \mathrm{Myr}$--$1 \mathrm{Gyr}$; in fact some of these systems transition from the thermal time-scale mass transfer mode into an RGB donor system.  Systems leaving behind a C/O or hybrid (a C/O core with a thick He mantle) WD have LMXB lifetimes of $1$-$100 \mathrm{Myr}$. 

Relevant modelling papers include \citep{king99,kolb00,tauris00,king01,podsiadlowski02,li02}.
\end{description}
\begin{description}
\item[CE $\rightarrow$ UC-LMXB] Once more massive donors reach the RGB, they have developed a convective envelope and their $\xi_{\mathrm{ad}} \lesssim 0 < \xi_L$.  When these donors start mass transfer, it is dynamically unstable, leading to a common envelope (CE) event \citep{paczynski76}.  In the CE, the orbital separation between the donor's core and the NS is reduced as orbital energy is tapped to expel the donor's envelope.  Systems in this region of initial phase space exit the CE in a tight enough orbit ($\Porb \lesssim 10 \mathrm{hr}$) that GW emission can drives the remnants back into contact within a Hubble time.  This initiates an ultracompact LMXB (UC-LMXB) phase of mass-transfer. The minimum $\Porb$ of such systems can be as short as $2$-$3\,\mathrm{minutes}$ and the general evolution trend is towards longer $\Porb$ as the (semi-)degenerate donor expands in response to mass loss.  The $\Mdot$ rate peaks at $\Porb$-minimum and decreases with increasing $\Porb$. The driver of mass-transfer is  Like the CV-like systems, UC-LMXBs have a mass-transfer lifetime that can persist for upwards of $10 \mathrm{Gyrs}$.

Relevant modeling papers include \citep{savonije86,podsiadlowski02,deloye03,nelson03,vandersluys05,deloye07}.
\end{description}
\begin{description}
\item[CE $\rightarrow$ Wide System] This region is similar to the CE systems leading to UC-LMXBs. Here though, the final post-CE separation is too wide for GW-emission to drive the binary back into contact and no LMXB phase of evolution occurs.
\end{description}

\begin{figure}
  \includegraphics[height=.425\textheight]{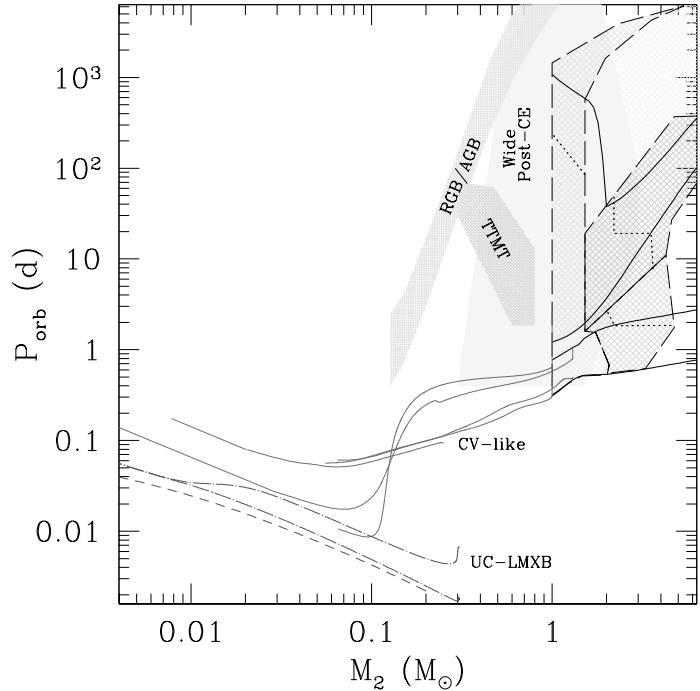}
  \caption{The evolution or endpoints for binaries starting mass transfer in each of the different regions shown in Fig \ref{fig:lmxb_ips}. The approximate endpoint phase-space for systems with a definitive $\Mdot$ turn-off point are shown by labelled regions: wide post-CE systems that do not undergo further phases of mass-transfer (light-gray shaded region, calculated using BSE \citep{hurley02} assuming a CE efficiency parameter of 1), post giant-branch donor systems (lighter gray, cross hatched region, calculated using BSE), and post thermal timescale mass-transfer systems (darker gray, cross hatched region, calculated using BSE).  The theoretically expected range of CV-like and ultracompact LMXB systems, which do not have a clear mass-transfer termination point, are shown with tracks.  Solid lines show the evolution of CV-like systems \citep[from][]{vandersluys05}, dash-dotted lines those of He-donor ultracompact systems \citep{deloye07}, and the dashed line the lower-bound for C-donor ultracompact systems \citep{deloye03}.
\ }
\label{fig:lmxb_endpoint}
\end{figure}

\begin{figure}
  \includegraphics[height=.425\textheight]{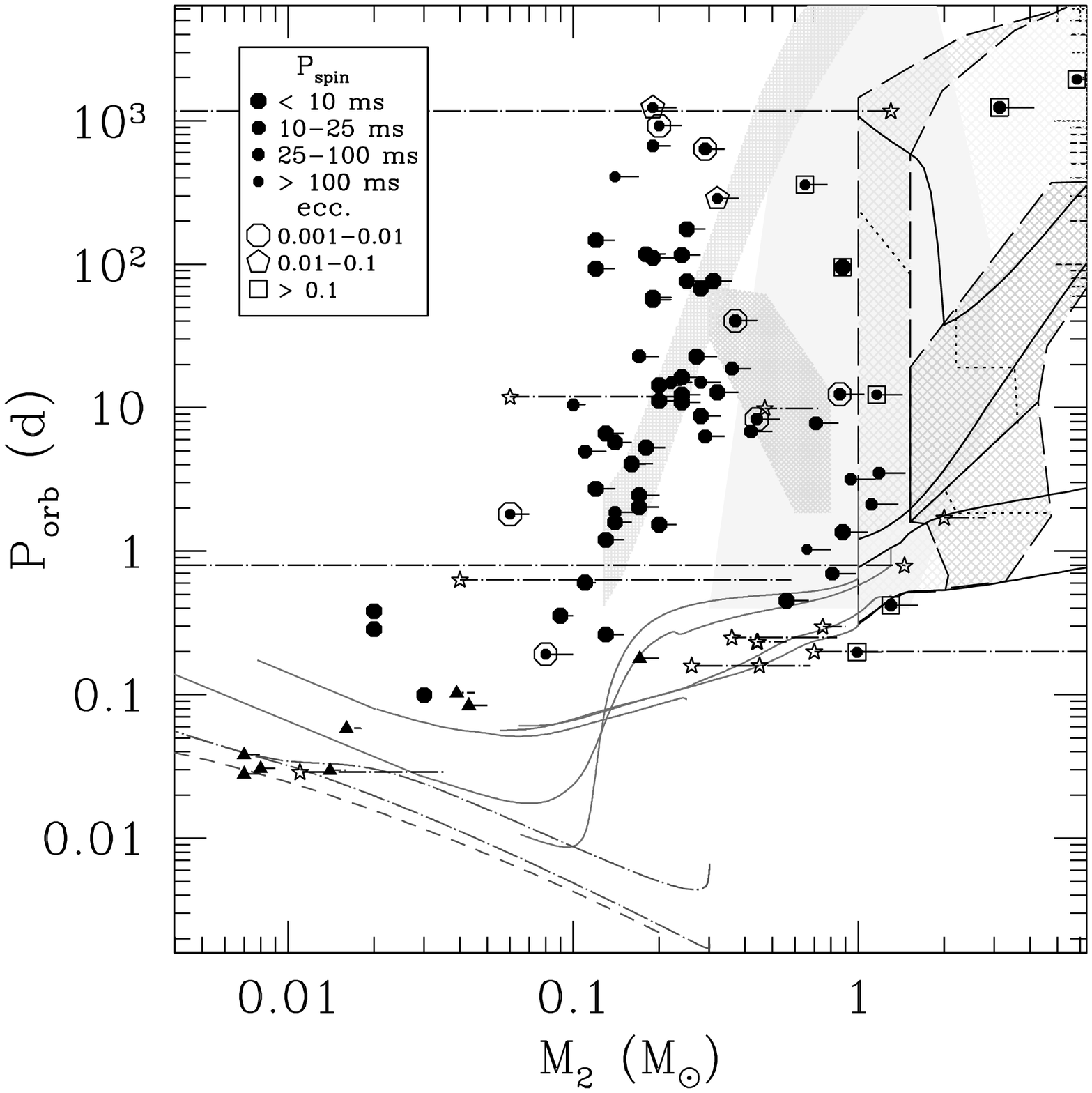}
  \caption{The comparison between LMXB-phase evolution or  end-states (i.e., Fig \ref{fig:lmxb_endpoint}  and the observed field bPSR and LMXB populations (i.e., Fig \ref{fig:bPSRs_sys} in the $\Porb$-$M_2$ plane.  There is a significant population of bPSRs systems generally consistent with being progeny of giant-branch or post thermal time-scale mass transfer systems at  $1 \mathrm{d} \lesssim \Porb \lesssim 100 \mathrm{d}$.  However, bPSRs with $\Porb \gtrsim 100 \mathrm{d}$ generally have $M_2$ smaller than predicted by theory and there are a cluster of MSPs between $0.1 \mathrm{d} \lesssim \Porb \lesssim 1 \mathrm{d}$ that are inconsistent with being progeny of any elucidated class of LMXBs. Other items of note are  the MSP system in a $e=0.4$orbit,  as well as a distinct lack of clearly identifiable bPSR progeny of any accreting MSP system. See text for further discussion.}
  \label{fig:sys_comp}
\end{figure}

Beyond differences in LMXB-phase lifetimes and evolution of $\Mdot$ rates, how mass transfer turn-off occurs also differs between the four classes of LMXBs discussed above. This is illustrated in Fig. \ref{fig:mdotcomp}, which shows representative examples of $\Mdot$ evolution for LMXBs from the RGB/AGB, thermal time-scale mass-transfer, and ultracompact classes.  The RGB/AGB and thermal time-scale systems both exhibit a rapid and clear termination in mass-transfer produced by the almost complete removal of the donor's envelope.  In UC-LMXB (as well as CV-like systems), mass-transfer never shuts-off.  Instead both these LMXB classes have gradually declining $\Mdot$ rates as the systems evolve to longer $\Porb$.  How this difference may affect the $\Pspin$-evolution will be discussed below.

Here, though, this distinction affects the phase space in which bPSR progeny of each LMXB class should be expected.  For RGB/AGB and thermal timescale systems, the answer is rather clear: their bPSR progeny should turn-on in the same region of phase space mass transfer turns-off.  These regions of $M_2$-$\Porb$ phase space are highlighted in Figure \ref{fig:lmxb_endpoint}.  The cross-hatched region labeled ``RGB/AGB'' shows the region giant-branch LMXBs are expected to leave their remnant bPSR systems.  This region shows the correlation between final $\Porb$ and remnant $M_2$ expected for this LMXB class as discussed above.  Additionally, thermal time-scale systems producing He WDs also have LMXB-phase endpoints on the lower half of this region.  The cross-hatched region labeled ``TTMT'' shows the endpoints for thermal time-scale systems leaving behind hybrid and C/O WD companions.  The solid, light-gray region labeled ``Wide Post-CE'' shows the endpoints of the post-CE systems not leading to UC-LMXBs. 

Since UC and CV-like LMXBs do not have a definitive $\Mdot$ turn-off point, identifying the phase space their bPSR progeny should occupy is not as clear cut.  Thus, in Fig \ref{fig:lmxb_endpoint}, I show the range of evolution in this $M_2$-$\Porb$ phase space expected for each class of LMXBs.  As discussed above, the tracks for CV-like systems (solid lines) initiating mass-transfer near the base of the RGB evolve to $\Porb$ values over-lapping ultracompact systems (dash-dotted and dashed lines).  However, the CV-like systems attaining the shortest $\Porb$ values take so long to evolve to there they do not have time to evolve back out to longer $\Porb$ \citep{vandersluys05}. In principle, any point along these tracks where the accretion disk is unstable and enters periods of quiescence could allow the pulsar to turn-on \citep{burderi01}.  However, as I'll discuss below, there is clear observational evidence for the LMXB phase persisting in systems with unstable disks, but no evidence as of yet for bPSRs detected in corresponding regions of phase space.

\section{Connecting LMXB and Pulsar Populations}
In Figure \ref{fig:sys_comp}, I overlay the observed bPSR and LMXB systems shown in Fig. \ref{fig:bPSRs_sys} with the evolutionary/end-point phases space available to the different classes of LMXBs as shown in Fig. \ref{fig:lmxb_endpoint}.  

Regarding the observed LMXBs in this plot, all have locations in this diagram consistent with that expected for at least one of the four LMXB-classes.  The only significant exceptions to this statement are the two accreting MSP systems with $\Porb \approx 2 \mathrm{hr}$, SAX J1808-365 and IGR 00291+593.  These systems lie in the general vicinity of the CV-like LMXB systems, but both have minimum-$M_2$ for their $\Porb$ that are larger than the upper-limits from standard evolution for this class.  This could be indicative that external irradiation is important to the donor's thermal evolution in these systems \citep{bildsten01}.  Also of note is the fact that all of the known accreting MSPs can be associated with either CV-like or ultracompact LMXBs.  Why this is the case is unclear, but could be connected to the lower $\Mdot$ values in these LMXB classes: accreting material at $\Mdot$-rates greater than a few percent $\Mdot_{\mathrm{Edd}}$ could screen the NS's magnetic field sufficiently to prevent magnetic channelling of the accretion flow \citep{cumming01}.

Amongst the bPSRs, there are several sub-populations whose location in the $M_2$-$\Porb$ plane are consistent theoretical expectations.  Those with minimum-$M_2 \gtrsim 0.1 \msun$ and $1.0 \lesssim \Porb \lesssim 20 \mathrm{d}$ are, generally speaking, consistent with being progeny of either  RGB-donor LMXBs or thermal time-scale mass transfer systems. Those that lie along the post-RGB $M_2$-$\Porb$ relation almost all have $\Pspin < 10 \mathrm{ms}$, consistent with expectations for NSs spun-up in a long-lived phase of sub-Eddington accretion.  Those systems in the region of post-thermal time-scale systems with C/O or hybrid WD companions have less-highly recycled NS.  Such systems experience a phase of super-Eddington $\Mdot$-rates and so are expected to host slower spinning NSs.  Non-eccentric, mildly recycled ($\Pspin > 10 \mathrm{ms}$) bPSRs in the vicinity of the post-CE region of this phase space are also consistent with being post-CE systems.  In this case, the NS would have been recycled via wind-accretion during the ABG phase preceding the CE event \citep{vandenheuval94}.  And, finally, systems with non-recycled NSs, $0.5 \lesssim M_2 \lesssim 1.4 \msun$ and $e > 0.1$ are likely systems in which the NS formed \emph{last}, its birth kick generating the system's larger eccentricity \citep[see][for discussion of this bPSR formation scenario]{tauris00,davies02,church06}.

This leaves several bPSR sub-populations that theory currently can not explain. The bPSRs with $M_2 \lesssim 0.5 \msun$ and  $\Porb \gtrsim 60 \mathrm{d}$ cluster towards donor masses too small to be consistent with theoretical expectations, as pointed out previously by \citet{tauris99}. Most bPSRs with $\Porb \lesssim 1 \mathrm{d}$ are not readily identifiable with any LMXB population. At this conference, Breton et al. presented a poster on one of these system, PSR J1744-3922 (the mildly eccentric $\Porb < 1 \mathrm{d}$ system with $M_2 \approx 0.1 \msun$ in Fig. \ref{fig:sys_comp}). There is one system in the group that may be associated with the CV-like LMXBs, the $\Porb \approx 0.1 \mathrm{d}$ system PSR J2051-0827, given its proximity to the two 2-hour accreting MSPs mentioned above.  If this is the case, this system might provide indirect evidence supporting the hypothesis that the pulsar in SAX J1808-365 turns-on in quiescence \citep[e.g.,][]{burderi03}.  Finally, also at this conference, D. Champion presented the discovery of PSR J1903+0327 ($\Porb = 95\,\mathrm{day}$, minimum-$M_2 = 0.88 \msun$) , a bona-fide MSP in a $e = 0.44$ orbit. The combination of a recycled pulsar in such an eccentric binary defies standard explanations since the mass transfer necessary to spin-up the pulsar should also damp-out the eccentricity. 

One other puzzling aspect of this diagram concerns bPSR progeny of the CV-like and ultracompact LMXBs: namely that there aren't any (apart from possibly PSR J2051-0827 discussed above). This is somewhat ironic given that the accreting MSPs---all of which appear to be members of these LMXB-classes---have been taken as the first, best direct evidence supporting the hypothesized LMXB-MSP connection.  Yet, most of the accreting MSPs appear not to have bPSR progeny---at least not any that have been detected.

\section{Some Open Questions}
So, do CV-like and ultracompact LMXBs (and the accreting MSP systems by extension) leave behind radio bPSRs? Possibly, no. This could be tied to the qualitative difference in mass-transfer shut off between these classes of LMXBs and the other two.  It has been pointed out \citep{jeffrey86,ruderman89} that a rapid shut-off of mass transfer is required to keep the NS spun up.  Rapid here means a the $\Mdot$ evolution time $\tau_{\Mdot} \equiv \Mdot/\ddot{M}$ that is shorter the NS spin-down time-scale $\tau_{\mathrm{spin}} \equiv \Pspin/\dot{P}_{\mathrm{spin}} \propto \Mdot^{-3/7}$ \cite{ruderman89}.  For GW-driven orbital evolution, in the limit $M_2 \ll M_{\mathrm{NS}}$ valid at late times,  $\tau_{\Mdot} \propto \Mdot^{-11/14}$.  Thus as $\Mdot$ decreases as the orbital separation grows, $\tau_{\mathrm{spin}}/\tau_{\Mdot}$ decreases.  For a NS with a magnetic field of $B = 5 \times 10^{8} \mathrm{G}$, these two timescales are roughly equal at an $\Mdot \approx 10^{-9} \msun \mathrm{yr}^{-1}$, so in post $\Porb$-minimum accreting MSPs, the systems have time to come into and maintain spin-equilibrium.  Thus the NS in these systems should spin-down as $\Mdot$ decreases, possibly preventing a radio pulsar from ever turning-on.

The measured $\Pspin$ in the ultracompact MSPs, however, argue that this simple picture is not be the whole story.  At $\Porb = 40 \mathrm{minutes}$, for example, the equilibrium-$\Pspin \sim 10$-$100 \mathrm{ms}$ (depending on $B \gtrsim 10^8 \mathrm{G}$ and donor properties setting $\Mdot$).  For the lowest $B$-field strengths and highest $\Mdot$, this range is barely consistent with the highest $\Pspin$ measured in the 4 ultracompact accreting MSPs at $\Porb > 40 \mathrm{minutes}$.  Thus it appears likely the NSs here are spinning faster than at equilibrium.  A plausible explanation for this is since these systems are transient, the NS accretes matter only during outbursts and thus at a rate significantly higher than the secular $\Mdot$. This could allow the NS to maintain a shorter $\Pspin$ even in the face of a declining secular $\Mdot$ \citep[see also,][]{li98}.

So, perhaps, NS spin-down is not the answer to why we do not see bPSR progeny of the accreting MSPs.  Work considering how the NS spin evolves during outburst and in the intervening quiescent phases to address this question in detail is indicated. The question of whether or the pulsar should ever turn-on in these systems where mass transfer never truly turns-off is a related, open question.  Another possibility is that current pulsar searches are not sensitive to finding bPRS in such compact orbits.  Having the observing community make quantitative statements concerning  detection limits for bPSRs searches in this portion of phase space is essential to determine whether we are indeed simply missing this population of bPSRs. 

Finally, binary evolution theory currently has no explanation for the populations of bPSRs with $\Porb \lesssim 1 \mathrm{d}$ or $\Porb \gtrsim 50 \mathrm{d}$, as discussed above.  Attempting to clarify the origin of these sub-populations is an obvious topic for future inquiry.

\begin{theacknowledgments}
I am happy to thank Ron Taam and Craig Heinke for helpful discussions during the preparation of this work.  I am very grateful to Marc van der Sluys for providing evolutionary tracks for CV-like systems and Jarrod Hurley for making publically available his BSE code.  This work was supported by NASA through the Chandra X-ray Center grant number TM7-8007X.
\end{theacknowledgments}


\bibliographystyle{aipproc}   


\bibliography{deloye_1}


\end{document}